\documentclass[aps,prl,reprint,twocolumn,superscriptaddress,floatfix,nofootinbib,longbibliography]{revtex4-1}
\usepackage{amsmath,amssymb,color,comment,physics}
\usepackage[makeroom]{cancel}
\usepackage[caption=false]{subfig}
\usepackage{mathrsfs}
\usepackage{graphicx}
\usepackage{subfig}
\usepackage[countmax]{subfloat}
\usepackage[english]{babel}
\usepackage[bookmarks=true,colorlinks,linkcolor=OrangeRed,urlcolor=NavyBlue,citecolor=RoyalBlue]{hyperref}
\usepackage[dvipsnames]{xcolor}
\usepackage{ulem} 
\usepackage{dsfont}

\definecolor{mygreen}{rgb}{0,0.5,0}
\definecolor{myblue}{rgb}{0,0,0.75}
\definecolor{mymagenta}{cmyk}{0,1,0,0.12}
\definecolor{mygray}{rgb}{0.5,0.5,0.5}

\definecolor{mypink1}{rgb}{0.858, 0.188, 0.478}
\definecolor{mypurple}{rgb}{0.49,0.18,0.56}
\definecolor{mygold}{rgb}{0.93,0.49,0.13}
\definecolor{mygreen}{rgb}{0,0.5,0}
\definecolor{myblue}{rgb}{0,0,0.75}
\definecolor{mymagenta}{cmyk}{0,1,0,0.12}
\definecolor{mygray}{rgb}{0.5,0.5,0.5}

\begin{document}

\title{Fate of Lattice Gauge Theories Under Decoherence}
\author{Jad C.~Halimeh}
\affiliation{INO-CNR BEC Center and Department of Physics, University of Trento, Via Sommarive 14, 38123 Povo (TN), Italy}

\author{Valentin Kasper}
\affiliation{ICFO - Institut de Ciencies Fotoniques, The Barcelona Institute of Science and Technology, Av.~Carl Friedrich Gauss 3, 08860 Castelldefels (Barcelona), Spain}
\affiliation{Department of Physics, Harvard University, Cambridge, MA, 02138, US}

\author{Philipp Hauke}
\affiliation{INO-CNR BEC Center and Department of Physics, University of Trento, Via Sommarive 14, 38123 Povo (TN), Italy}

\begin{abstract}
A major test of the capabilities of modern quantum simulators and NISQ devices is the reliable realization of gauge theories, which constitute a gold standard of implementational efficacy. In addition to unavoidable unitary errors, realistic experiments suffer from decoherence, which compromises gauge invariance and, therefore, the gauge theory itself. Here, we study the effect of decoherence on the quench dynamics of a lattice gauge theory. Rigorously identifying the gauge violation as a divergence measure in the gauge sectors, we find at short times that it first grows diffusively $\sim\gamma t$ due to decoherence at environment-coupling strength $\gamma$, before unitary errors at strength $\lambda$ dominate and the violation grows ballistically $\sim\lambda^2t^2$. We further introduce multiple quantum coherences in the context of gauge theories to quantify decoherence effects. Both experimentally accessible measures will be of independent interest beyond the immediate context of this work.
\end{abstract}

\date{\today}
\maketitle

Gauge theories are not only of fundamental importance to understanding vastly different states of matter ranging from particle physics to strongly correlated materials \cite{Weinberg_book,Wen2017}, they are also relevant for quantum information technologies \cite{Dennis2002,Wang2003}. Gauge invariance entails a local conservation law, which is a defining characteristic of any gauge theory. In condensed matter physics, gauge theories appear as effective descriptions of strongly correlated matter \cite{Hermele2004,Rokhsar1988, Moessner2001}, whereas in particle physics gauge invariance is usually postulated \cite{Peskin2016}. Despite the conceptual elegance of gauge symmetry, the numerical simulation of quantum many-body systems involving gauge fields is numerically extremely challenging, in and out of equilibrium \cite{Aoki2020, CarmenBanuls2020}. An alternative to explicit numerical calculation on classical computers is to use quantum simulators---devices specially designed to solve quantum many-body problems \cite{Zohar2011,Wiese2013,Dalmonte_review,Banuls_review,Alexeev2019}. The first experimental results for the quantum simulation of  gauge theories were obtained with trapped ion systems \cite{Martinez2016,Kokail2019}, followed by further experiments using ultracold atoms \cite{Bernien2017, Schweizer2019, Gorg2019, Mil2020, Yang2020} and superconducting qubits \cite{Klco2018,Klco2019}. Typically these implementations try to ensure unitary dynamics. However, in realistic situations the coupling to the environment cannot be completely suppressed. 
As such, the strong progress in quantum simulation renders it necessary to develop a better understanding of gauge-theory dynamics in the context of open systems. 

In this work, we study the dynamics of gauge theories in the presence of decoherence, focusing on a paradigmatic model that is characteristic of ongoing experimental efforts: an Abelian $\mathrm{Z}_2$ lattice gauge theory in one spatial dimension \cite{Schweizer2019}. In the presence of experimentally motivated unitary errors, starting from a gauge-invariant state we find a crossover from diffusive to ballistic spread of the quantum state over the gauge sectors at short times, which we quantify by rigorously relating the gauge violation to a divergence measure across different gauge-symmetry sectors in Hilbert space. Furthermore, we introduce a measure for the coherence of gauge violations by extending the multiple quantum coherences to gauge theories, which is shown to be essential for discerning coherent from incoherent gauge violations at late times.

Our work complements, from a quantum-simulation and far-from-equilibrium perspective, other investigations such as the study of dissipation in the context of topological quantum information, where a large body of literature exists on thermal heat baths \cite{Trebst2007,Hamma2009} as well as the dissipative preparation of topological ground states \cite{Weimer2010}. In the context of subatomic physics, open system dynamics has been employed, e.g., to study nuclear physics as in the melting of heavy quarkonium \cite{Akamatsu2018,Lehmann2020}, and techniques such as stochastic quantization exist to simulate unitary dynamics of gauge fields using nonunitary dynamics \cite{Damgaard1987}. From an orthogonal perspective, engineered dissipation has also been proposed to protect gauge invariance \cite{Stannigel2014,Lamm2020}.

\begin{figure}[htp]
	\centering
	\includegraphics[width=.4\textwidth]{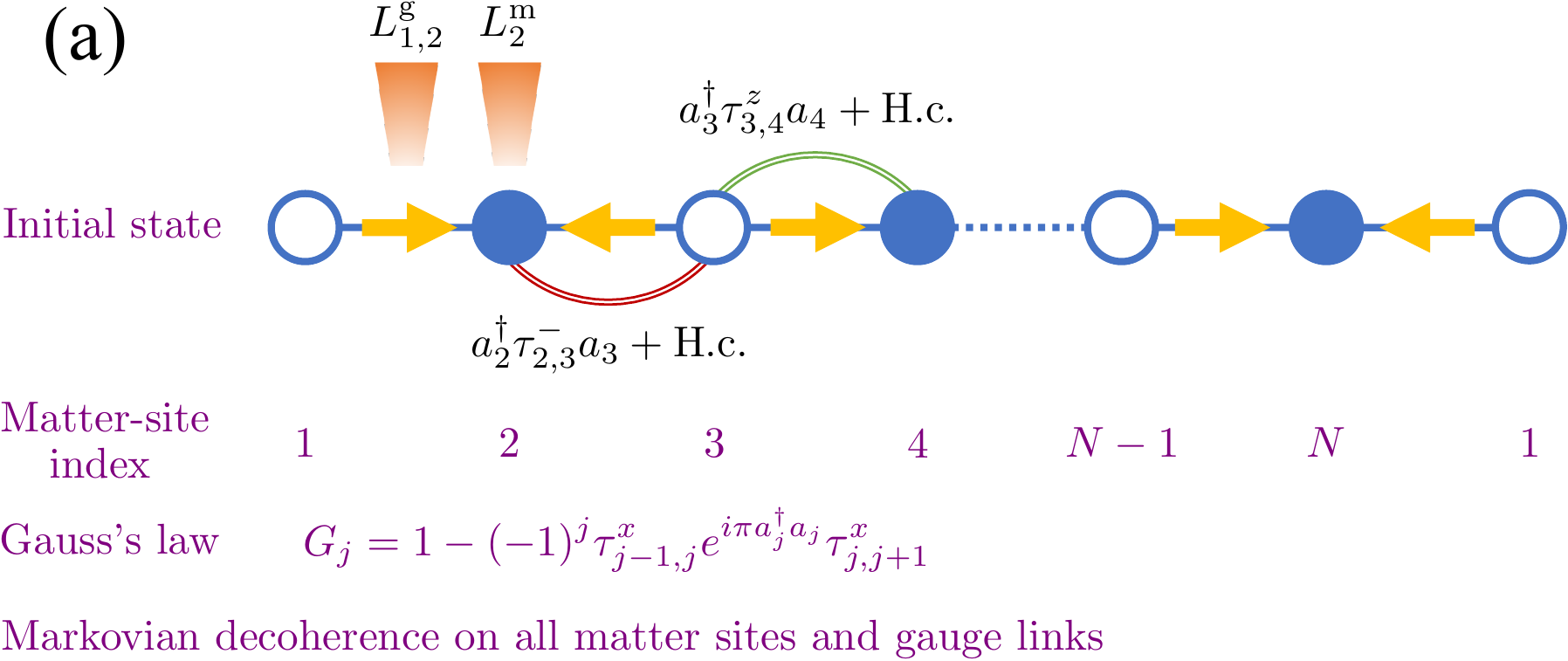}\\
	\vspace{0.15cm}
	\includegraphics[width=.48\textwidth]{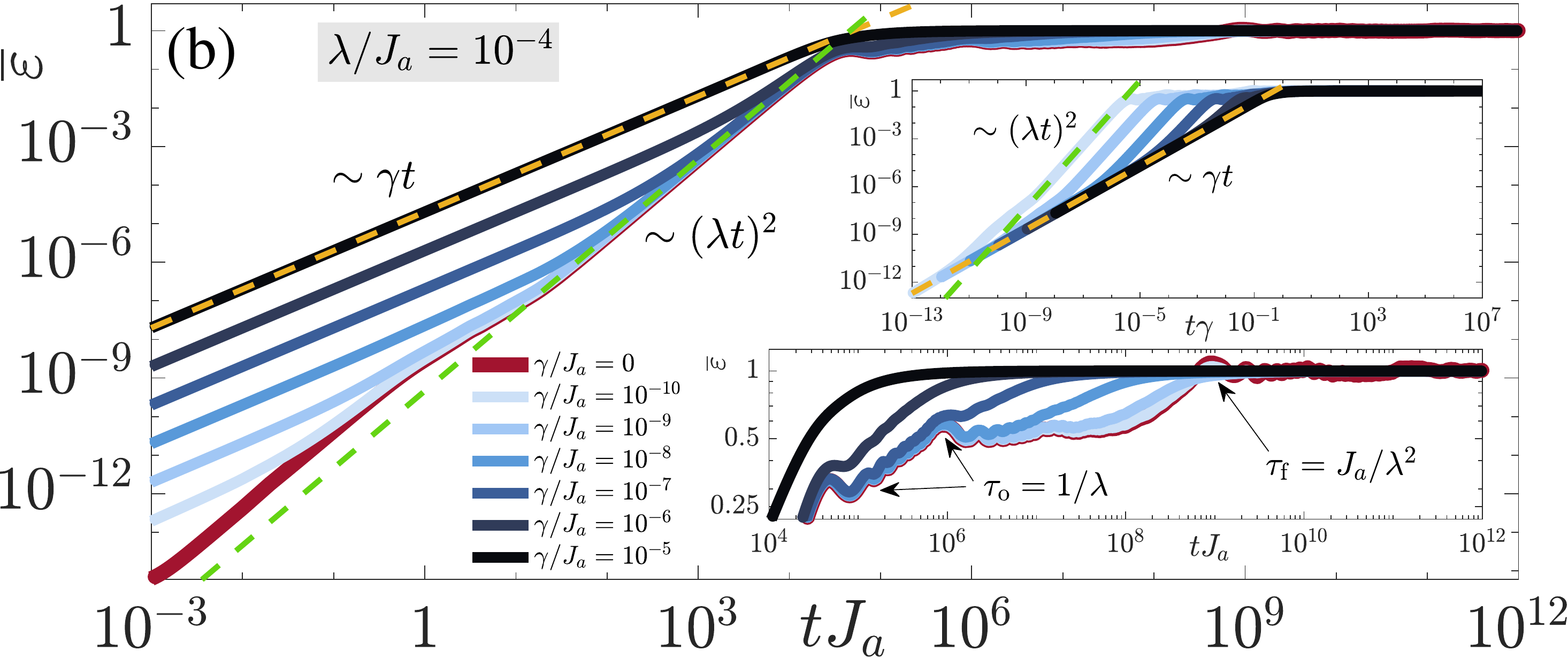}\\
	\includegraphics[width=.48\textwidth]{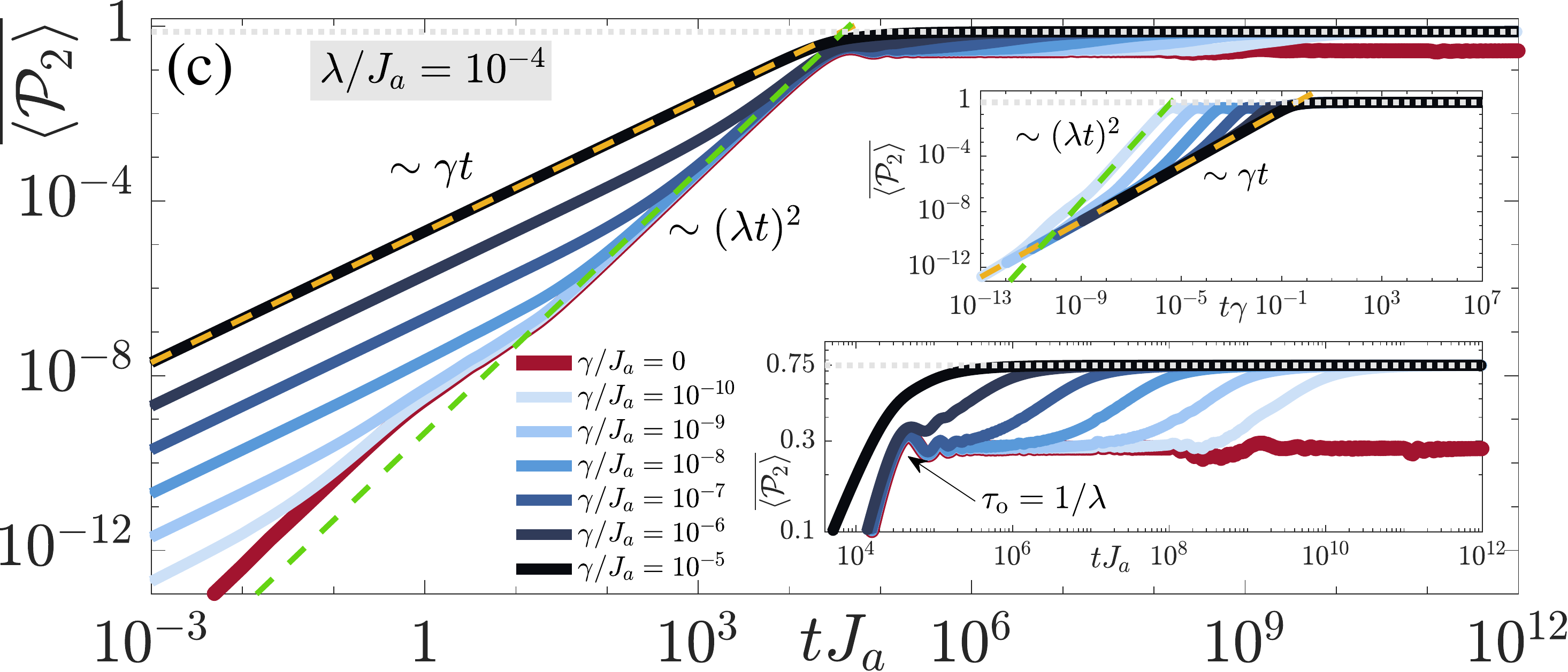}\\
	\includegraphics[width=.48\textwidth]{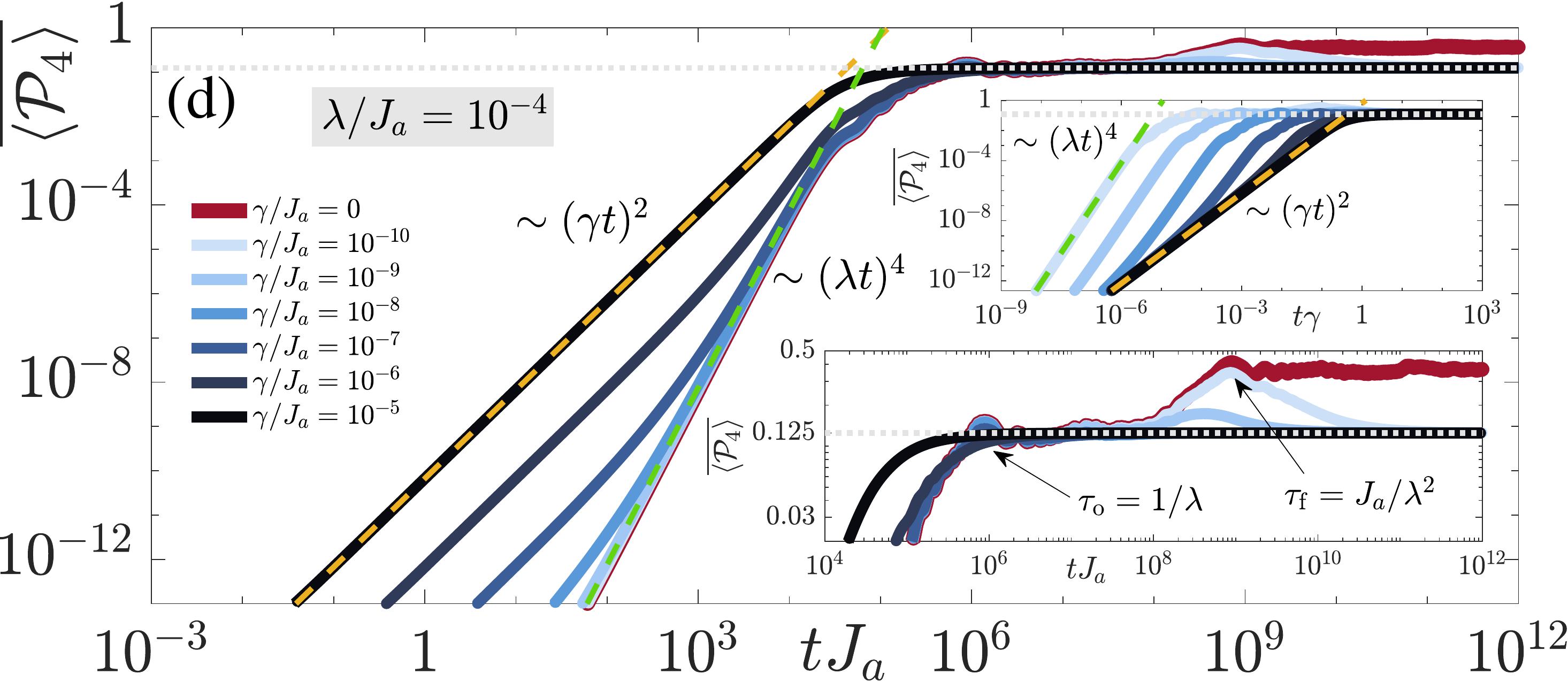}
	\caption{(Color online). Quench dynamics in a $\mathrm{Z}_2$ lattice gauge theory under gauge-breaking coherent and incoherent errors as are typical of state-of-the-art quantum-simulation experiments. 
		(a) The $\mathrm{Z}_2$ LGT, governed by Hamiltonian $H_0$. Gauge invariance between matter and gauge fields is embodied in conserved Gauss's-law generators $G_j$. Gauge invariance is violated due to coherent errors such as $a_j^\dagger \tau^-_{j,j+1} a_{j+1}+\mathrm{H.c.}$ composing the Hamiltonian $\lambda H_1$ as well as dephasing and dissipative effects with Lindblad operators $L^\mathrm{g,m}_j$.
		(b) Running average of the gauge violation at fixed $\lambda=10^{-4}J_a$ and for various values of the environment-coupling strength $\gamma$. 
		At $t\propto\gamma/\lambda^2$, a crossover occurs from diffusive behavior $\varepsilon\sim\gamma t$ due to incoherent errors to a ballistic regime $\varepsilon\sim(\lambda t)^2$ (see upper inset). The lower inset shows the effect of decoherence on the prethermal plateaus that in a closed system occur at timescales $\lambda^{-1}$ and $J_a\lambda^{-2}$. 
		(c) Running average of expectation value of supersector projector $\mathcal{P}_2$, showing the same scaling behavior as $\varepsilon$. 
		(d) Running average of the expectation value of (super)sector projector $\mathcal{P}_4$ with the higher-order diffusive $\sim\gamma^2t^2$ and ballistic $\sim\lambda^4t^4$ scalings, but the crossover occurs at the same timescale as in (b). At times $t\gtrsim1/\gamma$, the projector expectation values depart from their purely unitary steady-state values such that $\langle\mathcal{P}_2\rangle/\langle\mathcal{P}_4\rangle=6$, indicating equal distribution among all gauge sectors.
	}
	\label{fig:Fig1} 
\end{figure}

\textbf{\emph{Model.---}}Even though our results are more general \cite{JS}, for clarity, and inspired by a recent cold-atom experiment \cite{Schweizer2019}, we focus in our analysis on a $\mathrm{Z}_2$ lattice gauge theory (LGT) in one spatial dimension \cite{Zohar2017,Barbiero2019,Borla2019} given by the Hamiltonian
\begin{align}\label{eq:H0}
H_0=&\,\sum_{j=1}^N\big[J_a\big(a^\dagger_j\tau^z_{j,j+1}a_{j+1}+\text{H.c.}\big)-J_f\tau^x_{j,j+1}\big], 
\end{align}
see Fig.~\ref{fig:Fig1}(a). 
We assume periodic boundary conditions, yielding a total lattice length of $2N$ degrees of freedom ($N$ matter sites and $N$ links). The link variables located between the matter sites $j$ and $j+1$, $\tau_{j,j+1}^{x(z)}$, are represented by the $x$ ($z$) Pauli matrices. The matter fields are represented by hard-core bosons with the ladder operators $a_j,a_j^\dagger$ on matter site $j$ satisfying the canonical commutation relations $[a_j,a_l]=0$ and $[a_j,a_l^\dagger]=\delta_{j,l}(1-2a_j^\dagger a_j)$. The dynamics of the matter field couples to the $\mathrm{Z}_2$ gauge field with strength $J_a$, and the electric field has energy $J_f$. Throughout this work, we set $J_a=1$ and $J_f=0.54$, following Ref.~\cite{Schweizer2019}, although we have checked that our results are independent of this choice. 

Gauge invariance is defined by the conservation of the local $\mathrm{Z}_2$ symmetry, generated by the Gauss's-law operators
\begin{align}\label{eq:Gj}
G_j=1-(-1)^j\tau^x_{j-1,j}e^{i\pi a_j^\dagger a_j}\tau^x_{j,j+1},
\end{align}
with eigenvalues $g_j$ and where $[H_0,G_j]=[G_j,G_l]=0,\,\forall j,l$. Ideally, the gauge theory thus preserves the eigenvalues $\mathbf{g}(0)$ into which the system is initialized. 

Of particular interest (and concern) to modern quantum simulators of the noisy intermediate-scale quantum (NISQ) device era are inherent experimental imperfections. 
In the absence of unrealistic fine-tuning, these manifest themselves in two main ways: (i) unitary gauge invariance-breaking errors such as, e.g., density-induced tunneling, unassisted matter tunneling, or unassisted gauge flipping \cite{Halimeh2020a}, and (ii) decoherence in the form of, e.g., particle loss, dephasing due to laser-light scattering, driving-induced heating, or magnetic-field fluctuations. For concreteness, we formalize the unitary (or coherent) gauge invariance-breaking errors by the Hamiltonian
\begin{align}\nonumber
\lambda H_1=&\,\lambda \sum_{j=1}^N\Big[\big(c_1a_j^\dagger \tau^-_{j,j+1} a_{j+1}+c_2a_j^\dagger\tau^+_{j,j+1} a_{j+1}+\text{H.c.}\big)\\\label{eq:H1}
&+a_j^\dagger a_j\big(c_3\tau^z_{j,j+1}-c_4\tau^z_{j-1,j}\big)\Big],
\end{align}
inspired by the building-block experiment of Ref.~\cite{Schweizer2019}, where $\lambda$ is the error strength and $c_{1,\ldots,4}$ are real numbers dependent on a dimensionless driving parameter $\chi$, and normalized to sum to unity. In this work we set $\chi=1.84$, leading to $c_1=0.51$, $c_2=-0.49$, $c_3=0.77$, and $c_4=0.21$, although we have checked that other values for $\chi$ and forms of $H_1$ yield the same qualitative picture.

Moreover, we model dissipative errors using the Lindblad master equation
\cite{Breuer_book,Manzano2020}
\begin{subequations}
\begin{align}\label{eq:EOM}
\dot{\rho}=&-i[H_0+\lambda H_1,\rho]+\gamma\mathcal{L}\rho,\\\nonumber
\mathcal{L}\rho=&\sum_{j=1}^N\Big(L^\text{m}_j\rho L^{\text{m}\dagger}_j+L^\text{g}_{j,j+1}\rho L^{\text{g}\dagger}_{j,j+1}\\
&-\frac{1}{2}\big\{L^{\text{m}\dagger}_j L^\text{m}_j+L^{\text{g}\dagger }_{j,j+1}L^\text{g}_{j,j+1},\rho\big\}\Big).
\end{align}
\end{subequations}
Here, $\rho(t)$ is the density matrix of the system at evolution time $t$, and $L^\mathrm{m}_j$ ($L^\text{g}_{j,j+1}$) is the jump operator at matter site $j$ [gauge link $(j,j+1)$] with environment-coupling strength $\gamma$.

Deviations from ideal gauge invariance can be quantified by the gauge violation and supersector projector, respectively, 
\begin{align}
\varepsilon(t)&=\frac{1}{N}\Tr\Big\{\rho(t)\sum_j\big[G_j-g_j(0)\big]\Big\},\label{eq:Measure}\\
\langle\mathcal{P}_M(t)\rangle&=\Tr\big\{\rho(t)\mathcal{P}_M\big\},\,\,\,\mathcal{P}_M=\sum_{\mathbf{g};\,\sum_jg_j=2M}P_\mathbf{g}.
\end{align}
Here, $P_\mathbf{g}$ is the projector onto the gauge-invariant sector $\mathbf{g}=\{g_1,g_2,\ldots,g_N\}$, and the system is initialized in the sector $\mathbf{g}(0)$. Further, $M$ denotes the number of local violations $g_j\neq0$. Consequently, $\mathcal{P}_M$ is the gauge-invariant supersector composed of all sectors with $M$ violations.

We can give further meaning to the gauge violation by rigorously connecting it to a divergence measure in Hilbert space, which can also be related to the overlap of two states only differing by a gauge transformation \cite{JS}. To this end, we define the mean-square displacement across the gauge sectors, $D^{(2)}_{\mathbf{g}(0)}(t)=\sum_{\mathbf{g}} d^2(\mathbf{g},\mathbf{g}(0))\Tr\{\rho(t) P_\mathbf{g}\}$, where 	$d^2(\mathbf{g},\mathbf{g}(0))=[\mathbf{g}-\mathbf{g}(0)]^2$. Since the projector is block diagonal in the gauge sectors, we obtain $D^{(2)}_{\mathbf{g}(0)}(t)=\Tr\{\rho(t)\sum_j [G_j-g_j(0)]^2\}$, and for the considered $\mathrm{Z}_2$ gauge theory 	$D^{(2)}_{\mathbf{g}(0)}(t)=2N\varepsilon(t)$. Thus, the gauge violation acquires a mathematically rigorous meaning as a mean-square displacement across the gauge sectors.

\textbf{\emph{Quench dynamics.---}} First, we investigate the effect of decoherence on the quench dynamics in an LGT. We consider an initial state $\rho_0$ in the sector $\mathbf{g}(0)=\{0,0,0,0\}$, which has even (odd) matter sites filled (empty) and gauge fields that point from odd to even matter sites; see Fig.~\ref{fig:Fig1}(a). In order to allow for larger system sizes in our exact diagonalization (ED) numerics, we consider here only dephasing on matter sites, which conserves the total particle number (as do $H_0$ and $H_1$), permitting us to reliably simulate the dynamics of an open-system LGT with $N=4$ matter sites over long evolution times.

The initial state is quenched at $t=0$ by $H=H_0+\lambda H_1$, with dephasing at matter sites given by $L^\text{m}_j=a_j^\dagger a_j$ and dissipation on gauge links by $L^\text{g}_{j,j+1}=\tau^z_{j,j+1}$, both of which at strength $\gamma$. The ensuing dynamics of the gauge-invariance violation and supersector projectors are shown in terms of their running temporal averages $\overline A(t)=\int_0^tds\, A(s)/t$ in Fig.~\ref{fig:Fig1}. 

As displayed in Fig.~\ref{fig:Fig1}(b), the gauge violation scales as $\varepsilon\sim\gamma t$ at earliest times. At sufficiently large $\gamma\gtrsim\lambda$, this behavior persists until $t\approx1/\gamma$, beyond which the gauge violation reaches its maximal value. However, for $\gamma\lesssim\lambda$, at $t\propto\gamma/\lambda^2$ the gauge violation undergoes a crossover to a scaling as $\varepsilon\sim\lambda^2t^2$. These scalings can be explained through time-dependent perturbation theory (TDPT) \cite{JS}, where the terms $\propto\,\gamma t^2,\,\gamma\lambda t^2$ and all coherent terms linear in $\lambda$ are shown to vanish.

Exploiting the identification of the gauge violation with a mean-square displacement in Hilbert space, it becomes possible to interpret the crossover from $\varepsilon\sim\gamma t$ to $\varepsilon\sim\lambda^2t^2$ as a crossover from diffusive to ballistic spreading through the gauge sectors, driven by the competition of dissipative and coherent quantum dynamics. 
Interestingly, the temporal sequence is inverted with respect to cases such as the dynamics of excitations in quantum networks, where ballistic spreading dominates at short times followed by diffusive behavior at later times \cite{Maier2019}. This occurs because coherent errors cannot contribute to the gauge violation at an order lower than $\propto\lambda^2 t^2$ in the case of a gauge-invariant initial state, as rigorously derived in TDPT \cite{JS}, and hence the scaling $\varepsilon\sim\gamma t$ is uncontested for $t\lesssim\gamma/\lambda^2$. 

As has been established through ED and a Magnus expansion in Refs.~\cite{Halimeh2020b,Halimeh2020c}, the coherent breaking terms with strength $\lambda$ give rise to a staircase of prethermal plateaus at timescales $J_a^{s-1}\lambda^{-s}$ with $s=1,2,\ldots,N/2$, with the remarkable finding that the final plateau is delayed exponentially in system size. Since the dominant incoherent contribution to the gauge violation is $\propto\gamma t$, this means that a competition will arise between the timescale $t\propto1/\gamma$, at which the gauge violation reaches a maximal value due to decoherence, and the prethermal-plateau timescales. The later plateaus are affected more prominently by decoherence; see lower inset of Fig.~\ref{fig:Fig1}(b). Indeed, for sufficiently small $\gamma\lesssim\lambda^2/J_a$ the prethermal plateaus are not compromised, and the effects of decoherence are hard to discern at later times in the gauge violation. Instead, for $\lambda^2/J_a\lesssim\gamma\lesssim\lambda$, the second (and, for $N=4$ matter sites, the final) plateau is visibly affected by incoherent errors, while for $\gamma\gtrsim\lambda$, the prethermal plateaus vanish altogether, along with the diffusive-to-ballistic crossover. Hence, even though a crossover from diffusive to ballistic spread occurs at $t\propto\gamma/\lambda^2$ for $\gamma\lesssim\lambda$, after which coherent errors are dominant, the contribution of decoherence to the gauge violation will dominate again at later times for sufficiently large $\gamma$. 

Interestingly, this behavior replicates itself in the projectors $\mathcal{P}_2$ and $\mathcal{P}_4=P_{\{2,2,2,2\}}$, whose dynamics are shown in Fig.~\ref{fig:Fig1}(c,d), respectively. The former behaves at short times very similarly to $\varepsilon$; however, the latter scales as $\langle\mathcal{P}_4\rangle\sim\gamma^2t^2$ in the diffusive and $\sim\lambda^4t^4$ in the ballistic regime, with the crossover time again being $t\propto\gamma/\lambda^2$ for $\gamma\lesssim\lambda$.
These scalings can also be explained in TDPT, as the gauge-breaking errors connect the supersectors $M=0$ and $M=4$ only at $\gamma^2$ and $\lambda^4$ as leading orders. Another interesting aspect in the dynamics of $\mathcal{P}_2$ and $\mathcal{P}_4$ is that at $t\gtrsim1/\gamma$, they depart from their purely unitary ($\gamma=0$) steady-state values such that $\langle\mathcal{P}_2\rangle\approx0.75$, $\langle\mathcal{P}_4\rangle\approx0.125$, and $\langle\mathcal{P}_0\rangle=1-\langle\mathcal{P}_2\rangle-\langle\mathcal{P}_4\rangle\approx0.125$. This leads to $\langle\mathcal{P}_2\rangle/\langle\mathcal{P}_4\rangle=\langle\mathcal{P}_2\rangle/\langle\mathcal{P}_0\rangle=6$, which is equal to the ratio of the number of gauge sectors in the associated supersectors. This behavior signals that the state has diffused equally across all allowed gauge sectors, in contrast to what happens under purely coherent gauge breakings, where interference effects prevent such an equal distribution \cite{Halimeh2020b,Halimeh2020c}. Here we note that $\langle\mathcal{P}_1\rangle=\langle\mathcal{P}_3\rangle=0$ at all times, because the Hilbert subspace of the system involves only even violations.

\textbf{\emph{Multiple quantum coherences.---}}To measure the effect of decoherence and distinguish it from that of coherent errors, we consider the multiple quantum coherence (MQC) spectra. MQCs are experimentally accessible quantities with a long history in nuclear magnetic resonance systems where they have been used to quantify quantum effects in many-body systems \cite{Pines1985, Pines1986, Pines1987}. Recently, the MQC protocol has also been exploited to study the decoherence of correlated spins \cite{SpinDecoherence2014}, many-body localization \cite{NMRLocalizationPRL2010, NMRLocalizationScience2015}, and as a witness for multiparticle entanglement in quantum many-body systems \cite{Gaerttner2018}. We will use MQCs to elucidate the connection between incoherent dynamics and local gauge invariance. 

Traditionally, MQCs have been used for separable high-temperature states, but recent experimental developments now allow for the study of pure entangled states in different platforms as for example in trapped-ion experiments \cite{Garttner2017}. In order to introduce the MQC, we start from the overlap measure $F_{\boldsymbol{\phi}}(t) \equiv \text{Tr}\left[\rho(t) \rho_{\boldsymbol{\phi}}(t) \right]$
between the density matrix of interest, $\rho(t)$, and a transformed version, $\rho_{\boldsymbol{\phi}}(t) \equiv U(\boldsymbol{\phi}) \rho(t) U^{\dagger}(\boldsymbol{\phi})$. For our purposes, $U(\boldsymbol{\phi}) = \prod_{i=1}^N e^{i\phi_j G_j}$, where $G_j$ are operators with a discrete spectrum and we generalize $\boldsymbol{\phi} = (\phi_1, \ldots, \phi_N)$ to be a real vector. 
In our case, the $G_j$ are the Gauss's-law operators, though construction of the MQC can be applied to any other set of commuting operators $G_j$ with discrete and bounded spectra. The density matrix can be decomposed as $\rho = \sum_{\Delta \mathbf{g}} \rho_{\Delta \mathbf{g}}$, with the coherences $\rho_{\Delta \mathbf{g}} (t) = \sum_{\mathbf{g}} P_{\mathbf{g}+\Delta \mathbf{g}} \, \rho \, P_{\mathbf{g}}$, and where $\Delta\mathbf{g}$ is the difference between the eigenvalues of gauge sectors $\mathbf{g}$ and $\mathbf{g}+\Delta \mathbf{g}$. The spectral decomposition of the overlap measure $F_{\boldsymbol{\phi}}(t)$ is given by 
\begin{align}
&F_{\boldsymbol{\phi}}(t) =  \sum_{\Delta \mathbf{g}} I_{\Delta \mathbf{g}}\left[{\rho}(t)\right] e^{-i \sum_j\phi_j\Delta g_j},
\end{align}
with the amplitudes $I_{\Delta \mathbf{g}}[\rho(t)]= \Tr [ \rho_{\Delta \mathbf{g}} (t) \rho_{-\Delta \mathbf{g}}(t)]$, called MQC intensities.

\begin{figure}[htp]
	\centering
	\includegraphics[width=.43\textwidth]{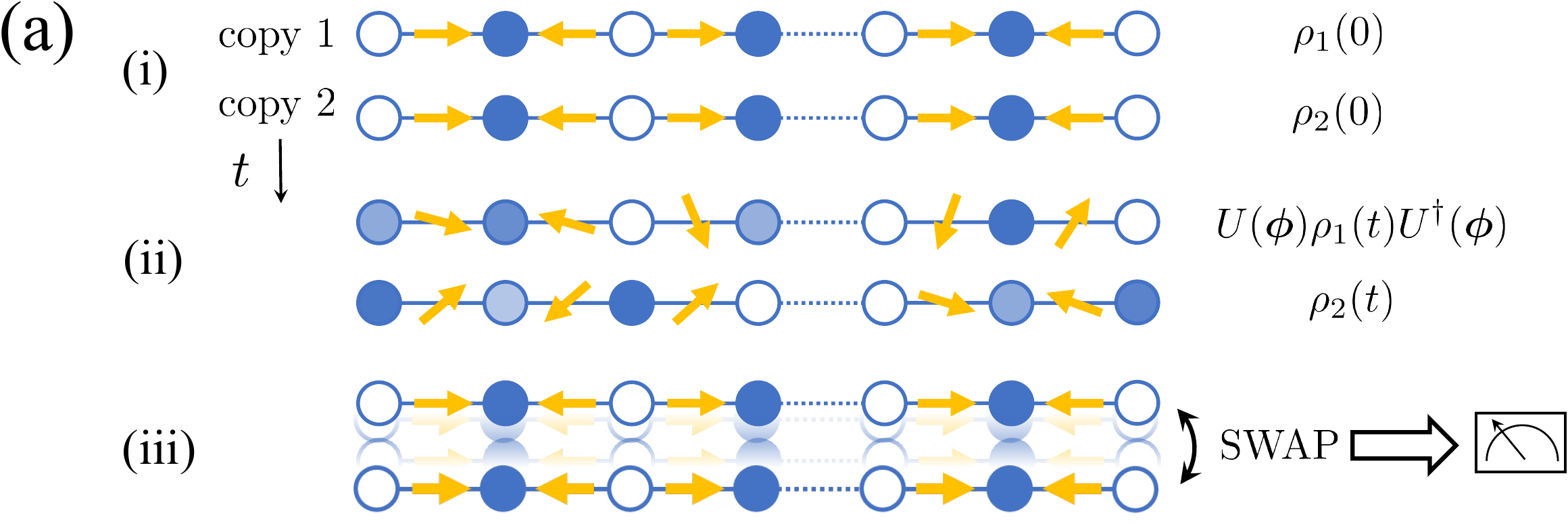}
	\includegraphics[width=.48\textwidth]{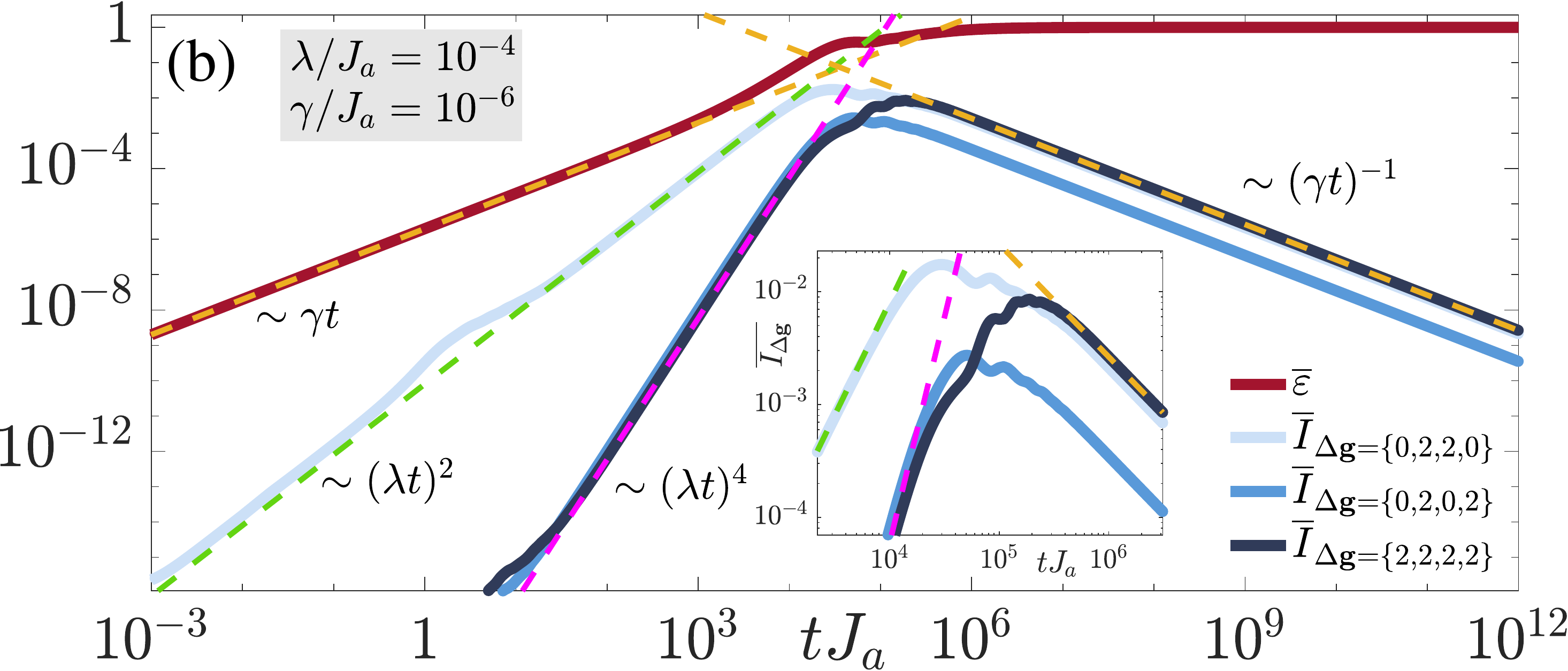}
	\begin{minipage}[c]{0.32\linewidth}
		\includegraphics[width=\linewidth]{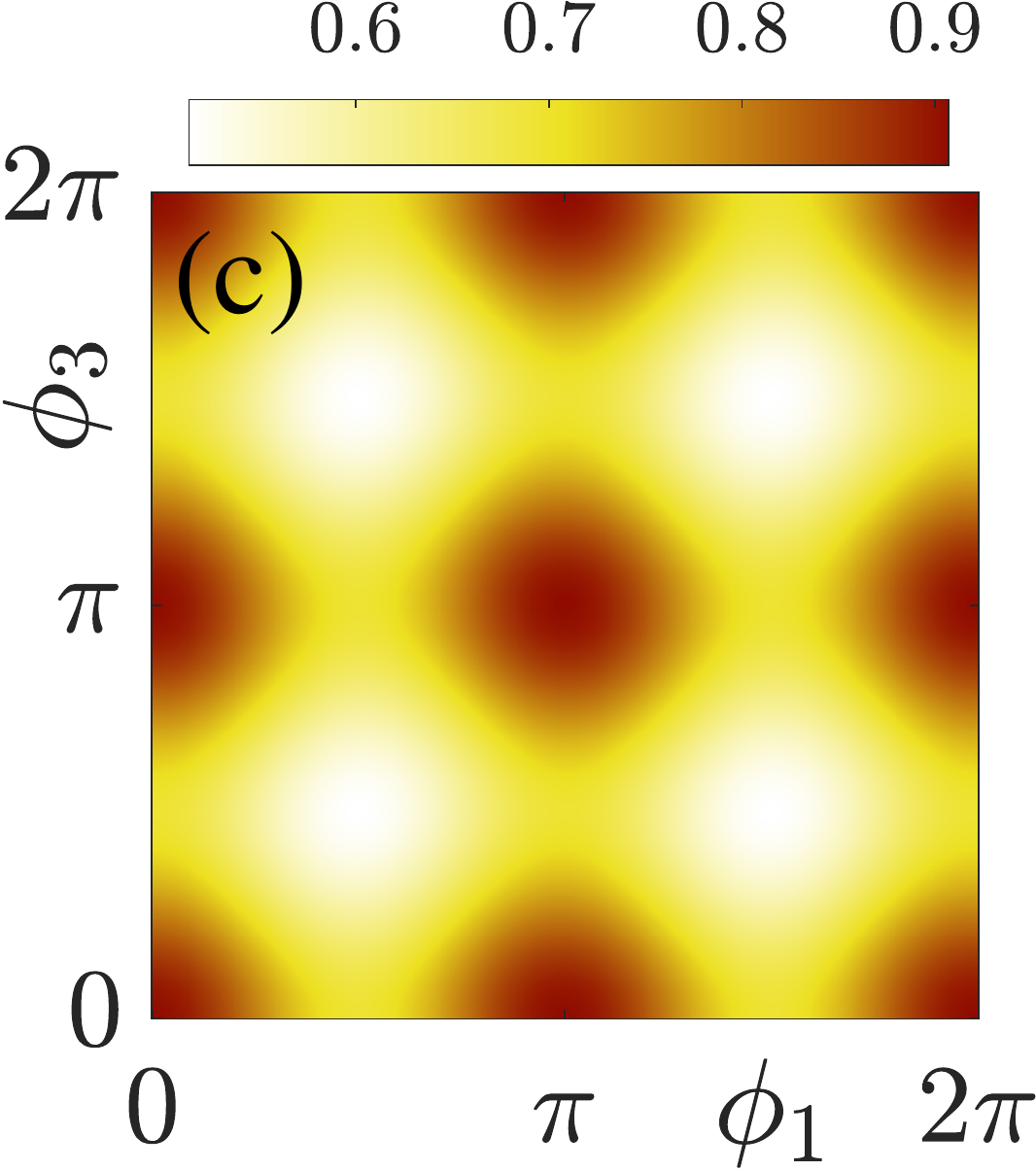}
	\end{minipage}
	\begin{minipage}[c]{0.32\linewidth}
		\includegraphics[width=\linewidth]{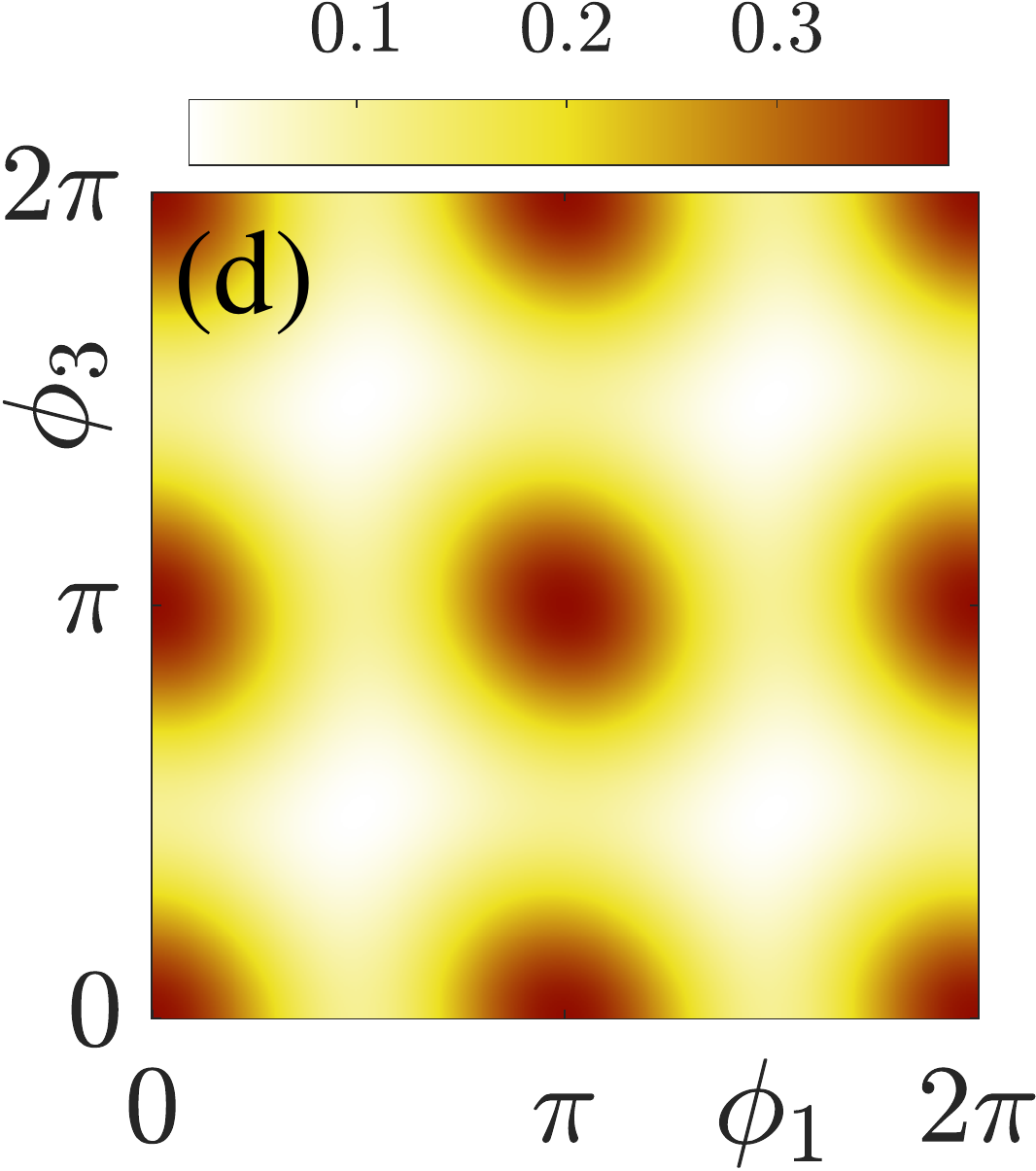}
	\end{minipage}
	\begin{minipage}[c]{0.32\linewidth}
		\includegraphics[width=\linewidth]{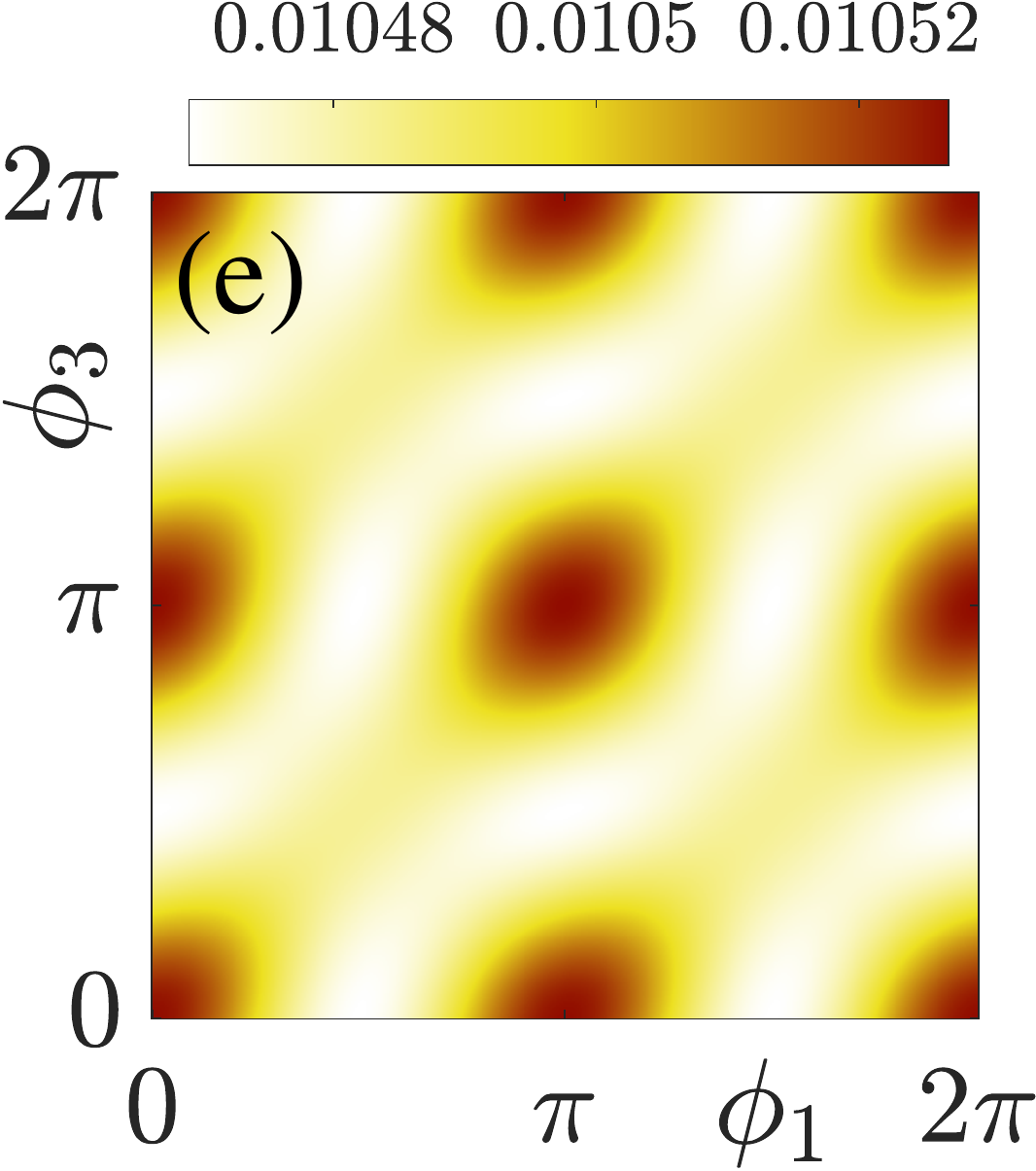}
	\end{minipage}
	\caption{(Color online). (a) Measurement protocol: (i) Prepare two independent copies of a quantum simulator of the $\mathrm{Z}_2$ lattice gauge theory. Allow both systems to evolve for a time $t$. (ii) Perform local unitary gauge transformations $U(\boldsymbol{\phi})$. This can be achieved by appropriately chosen laser fields coupled to the Gauss’s-law operators. (iii) Perform the SWAP operation, which exchanges the two systems, and also local measurements of the occupation and spin of the sites and the links respectively, thereby allowing the determination of $F(\boldsymbol{\phi})$. (b) Multiple quantum coherences for three dominant gauge breakings at fixed strengths of coherent and incoherent errors of $\lambda=10^{-4}J_a$ and $\gamma=10^{-6}J_a$, respectively. For reference, we also plot the associated gauge violation. (c-e) Fourier transform $F(\phi_1,0,\phi_3,0)$ of the multiple quantum coherences at evolution times $t=10^4/J_a$ (c), $t=10^5/J_a$ (d), and $t=10^6/J_a$ (e).}
	\label{fig:Fig2} 
\end{figure}

The MQC is also experimentally accessible in setups such as of Ref.~\cite{Kaufman2016} in the following way; see Fig.~\ref{fig:Fig2}(a). Initially, one prepares two copies of the state $\rho(0)$, i.e., $\rho(0) \otimes \rho(0)$, followed by a time evolution leading to $\rho(t) \otimes \rho(t)$. Next, one applies the rotation 
$U(\boldsymbol{\phi})$ on the first copy: $U(\boldsymbol{\phi}) \rho(t )U^{\dagger}(\boldsymbol{\phi}) \otimes \rho(t)$. Finally, one measures the expectation value of the swap operator $\mathds{S}$, which is
defined by $ \mathds{S}( \ket{\alpha} \otimes \ket{\beta}) = \ket{\beta} \otimes \ket{\alpha}$. Using this 
definition, one can show that the expectation value of the swap operator gives the MQCs: $\Tr [\mathds{S} \big(U(\boldsymbol{\phi}) \rho(t )U^{\dagger}(\boldsymbol{\phi}) \otimes \rho(t)\big)] = F_{\boldsymbol{\phi}}(t) $. 
The MQC intensities can then be extracted by Fourier transform with respect to $\boldsymbol\phi$.

Numerical results for both the MQC intensity and spectrum in the quench dynamics of the $\mathrm{Z}_2$ gauge theory are shown in Fig.~\ref{fig:Fig2}. In agreement with TDPT \cite{JS} at early times $I_{\{0,2,2,0\}}\sim(\lambda t)^2$, since the associated breaking with respect to $\mathbf{g}(0)=\{0,0,0,0\}$ is a first-order process in $H_1$, while $I_{\{0,2,0,2\}},I_{\{2,2,2,2\}}\sim(\lambda t)^4$ as the associated breakings correspond to second-order processes in $H_1$. In case of no decoherence, these intensities would reach a steady maximal value at sufficiently long times \cite{JS}. In the presence of decoherence, however, their running temporal averages begin to decay $\sim1/(\gamma t)$ at $t\approx1/\gamma$ when decoherence becomes the dominant process. Figure~\ref{fig:Fig2}(c-e) displays MQC spectra $F(\phi_1,0,\phi_3,0)$ at evolution times $tJ_a=10^4$, $10^5$, and $10^6$. Unlike the unitary case, a finite decoherence strength $\gamma$ suppresses the spectrum magnitude over time. 
Thus, the MQC provides a clear way to quantify how coherences between different gauge sectors, which are built up due to unitary processes, diminish as a direct result of dissipation and dephasing processes, even when the gauge violation itself may not show any discernable difference between the cases of zero and finite yet small $\gamma$; cf.~Fig.~\ref{fig:Fig1}(b). 

\textbf{\emph{Conclusion.---}}We have investigated the effects of decoherence in the form of dissipation and dephasing on the quench dynamics in lattice gauge theories. We have rigorously related the gauge violation to a divergence measure, enabling us to identify the domination of gauge violation by incoherent or coherent sources with a crossover from diffusive to ballistic spread throughout gauge sectors. 	
Another characteristic difference between coherent wave-like dynamics and diffusive evolution is the final distribution across gauge sectors: for dissipative errors, it becomes fully equal while interference effects prevent this for coherent errors. 
Even though at early to intermediate times one can distinguish between the dominance of incoherent and coherent contributions in the gauge violation from the associated perturbative scalings, at long times this is not possible. To achieve this, we have computed multiple quantum coherences as a reliable way to quantify gauge violations due to decoherence.
Our findings also carry a positive message for ongoing gauge-theory quantum simulations: as a consequence of the slow perturbative increase of gauge violations, there exists a well-defined short-time window within which local observables reliably reproduce the ideal gauge-theory dynamics.

\medskip

This work is part of and supported by the DFG Collaborative Research Centre SFB 1225 (ISOQUANT), the Provincia Autonoma di Trento, Q@TN, and the ERC Starting Grant StrEnQTh (Project-ID 804305). 
V.K.\ acknowledges support from ERC AdG NOQIA, Spanish Ministry of Economy and Competitiveness (Severo Ochoa program for Centres of Excellence in R\&D (CEX2019-000910-S), Plan National FISICATEAMO and FIDEUA PID2019-106901GB-I00\/10.13039 / 501100011033, FPI), Fundacio Privada Cellex, Fundacio Mir-Puig, and from Generalitat de Catalunya (AGAUR, 2017 SGR 1341, CERCA program, QuantumCAT U16-011424, co-funded by ERDF Operational Program of Catalonia 2014-2020), MINECO-EU QUANTERA MAQS (State Research Agency (AEI) PCI2019-111828-2 /10.13039/501100011033), EU Horizon 2020 FET-OPEN OPTOLogic (899794), the National Science Centre, Poland-Symfonia 2016/20/W/ST4/00314, and the European Union under Horizon2020 (PROBIST 754510). 

\bibliography{DecoherenceBiblio}
\end{document}